\documentclass[
aps,%
12pt,%
final,%
notitlepage,%
oneside,%
onecolumn,%
nobibnotes,%
nofootinbib,%
superscriptaddress,%
noshowpacs,%
centertags]%
{revtex4}

\begin{document}
\selectlanguage{english}

\title{Detection of Giant Pulses from the Pulsar PSR B0031--07}

\author{\firstname{A.~D.}~\surname{Kuzmin}}
\email{akuzmin@prao.psn.ru}
\affiliation{
Pushchino Radio Astronomy Observatory, Astrospace Center, Lebedev Physical Institute,
}
\author{\firstname{A.~A.}~\surname{Ershov}}
\email{ershov@prao.psn.ru} \affiliation{ Pushchino Radio Astronomy
Observatory, Astrospace Center, Lebedev Physical Institute,
}
\author{\firstname{B.~Ja.}~\surname{Losovsky}}
\email{blos@prao.psn.ru} \affiliation{ Pushchino Radio Astronomy
Observatory, Astrospace Center, Lebedev Physical Institute, }


\begin{abstract}
Giant pulses have been detected from the pulsar PSR B0031--07. A
pulse with an intensity higher than the intensity of the average
pulse by a factor of 50 or more is encountered approximately once
per 300 observed periods. The peak flux density of the strongest
pulse is 530 Jy, which is a factor of 120 higher than the peak
flux density of the average pulse. The giant pulses are a factor
of 20 narrower than the integrated profile and are clustered about
its center.

Key words: pulsars and black holes; neutron stars, giant pulses,
PSR B0031--07.
\end{abstract}

\maketitle

\section{Introduction}
Giant pulses (GPs), which are short-duration outbursts of pulsar
radio emission, are rare events that have been observed only in
five pulsars: the Crab pulsar PSR B0531+21 (Staelin and Sutton
1970), the millisecond pulsars PSR B1937+21 (Wolszczan et al.
1984) and PSR B1821--24 (Romani and Johnston 2001), the pulsar PSR
B1112+50 (Ershov and Kuzmin 2003), and the extragalactic pulsar in
the Large Magellanic Cloud PSR B0540--69 (Johnston and Romani
2003).

For normal pulsars, the intensity of their individual pulses
exceeds that of the average pulse by no more than several times.
The GPs from the Crab pulsar PSR B0531+21 (Kostyuk et al. 2003),
the extragalactic pulsar PSR B0540--69 in the Large Magellanic
Cloud (Johnston and Romani 2003), and the millisecond pulsar PSR
B1937+21 (Wolszczan et al. 1984; Cognard et al. 1996) exceed in
intensity the average level by factors of 70 000, 5000, and 300,
respectively; for the pulsars PSR B1112+50 and PSR B1821--24, this
excess reaches a factor of 80 (Ershov and Kuzmin 2003; Romani and
Johnston 2001).

Kuzmin and Losovsky (2002), Kostyuk et al. (2003), and Hankins et
al. (2003) have found extremely high brightness temperatures, $T_B
\sim 10^{35} \div 10^{37}$ K for the GPs from the millisecond
pulsar PSR B1937+21 and the Crab pulsar PSR B0531+21.

The characteristic features of the GPs also include their short
duration, stable position, and the pattern of their intensity
distribution. The GP duration is much shorter than that of the
integrated profiles for these pulsars. The GPs are localized
within a small part of the integrated profile. The intensity has a
two component distribution: lognormal for most of the pulses and a
power law $N\propto S^\alpha$ for pulses with an intensity higher
than a certain level, i.e., for GPs (Lundgren et al. 1995). The
boundary at which the pattern of the distribution changes
corresponds to an approximately 30-fold intensity of the average
pulse.

Attempts to detect GPs from other pulsars (Phinney and Taylor
1979; Johnston and Romani 2002; McLaughlin and Cordes 2003) have
failed so far.

We have detected GPs from the pulsar PSR B0031--07.

\section{OBSERVATIONS AND DATA REDUCTION}
The observations were carried out from April 19 through August 10,
2003, with the BSA radio telescope at the Pushchino Radio
Astronomy Observatory (Astrospace Center of the Lebedev Institute
of Physics, Russian Academy of Sciences). Only linear polarization
was received. We used a 128-channel receiver with the channel
bandwidth $\Delta f = 20$ kHz. The frequency of the first
(highest-frequency) channel was 111.870 MHz. The sampling interval
was 0.819 ms, and the time constant $\tau = 1$ ms. The
observations were performed in the mode of recording individual
pulses. The duration of one observing session was about 3 min. We
observed 205 pulsar periods in a session.

Over the above time interval, we conducted a total of 56 observing
sessions containing 11 480 pulsar periods. We detected 43 pulses
with peak flux densities exceeding the peak flux density of the
average pulse (for all of the 56 days of observations) by more
than a factor of 50 (one pulse per 270 observed periods). A sample
record of a giant pulse is shown in Fig. 1.

Figure 2 shows the strongest observed GP in comparison with the
integrated pulsar profile for all of the days of observations. The
peak flux density of this GP is a factor of 120 higher than that
of the average pulse. For a convenient examination, the integrated
profile was magnified by a factor of 100, and the flux densities
of the GP and the integrated profile are indicated on the left-
and right-hand scales of the vertical axes. The measured GP
duration is $w_{GP50} = 5 \pm 1$ ms. The duration of the integrated
pulsar profile is $w_{50} = 110$ ms $= 115$ mP $= 42^\circ$ (at
half maximum\footnote{The pulse duration was determined at half
maximum for each component of the two-component pulsar profile.
}). Thus, the GP is approximately a factor of 20 narrower than the
integrated profile, which is also typical of GPs.

This figure also shows the phases of the 43 observed GPs whose
peak flux density exceeds that of the average pulse by more than a
factor of 50. The temporal positions of the GPs are stable; they
are clustered about the middle of the first component of the
integrated profile. The phase difference between the GP centroid
and the integrated profile is $\Phi_{GP} - \Phi_{Av} = - 6^\circ
\pm 4^\circ = ( - 0.12 \pm 0.09 )\times w_{50}$.

We determined the
GP flux density by a comparison with the average pulse of this
pulsar with a known (period-averaged) flux density S. At a close
frequency of 102 MHz, the flux density of the average pulse is $S
= 400$ mJy (Izvekova et al. 1981); for the spectral index $\alpha
= -1.75$, this value corresponds to $S = 350$ mJy at the frequency
of our measurements, 111 MHz. The peak flux density of the average
pulse is

$$S_{Av}^{max} = S / k_{shape} = 4.4, Jy,$$

where $S$ is the period-averaged flux density of the average pulse,
and $k_{shape} = 0.08$ is the scaling ratio of the period-averaged
flux density of the pulsar to the peak flux density of the pulse
with allowance made for the shape of the integrated profile. The
GP peak flux density is

$$S_{GP}^{max} = S_{Av}^{max} \times(I_{GP}/I_{Av}),$$

where $I_{GP}/I_{Av}$ is the observed intensity ratio of the giant
pulse and the integrated profile. For the strongest GP,
$I_{GP}/I_{Av} = 120$, which corresponds to the peak flux density
$S_{GP}^{max} = 530$ Jy.

The GP brightness temperature is

$$T_{B} = S\lambda ^2/2k\Omega,$$

where $\lambda$ is the wavelength of the received radio emission,
$k$ is the Boltzmann constant, and $\Omega$ is the solid angle of
the radio-emitting region. Assuming the size of the radio-emitting
region to be $l \leq c\times w_{GP50}$, where $c$ is the speed of
light, and the pulsar distance to be $d = 0.68$ kpc (Taylor et al.
1995), for $w_{GP50} = 5$ ms and $S = 530$ Jy we obtain $T_{B}
\geq 10^{26}$ K.

The GPs that we detected are not the result of
scintillations. The characteristic scintillation timescale in the
interstellar medium (the radius of the time correlations) for a
pulsar with a dispersion measure $DM$ = 10.89 pc$~cm^{-3}$ at
the frequency of our observations, 111 MHz, is $\approx$ 1 min and
$\approx$ 1000 d, respectively, for refractive and diffractive
scintillations (Shishov et al. 1995). This time scale signi.cantly
exceeds the lifetimes of the observed GPs (no more than a few
seconds).

\section{DISCUSSION}
The GPs that we detected from the pulsar PSR B0031--07 exhibit
characteristic features of the classical giant pulses from the
pulsars PSR B0531+21 and PSR B1937+21. The GP peak intensity
exceeds that of the average pulse by more than a factor of 100.
The GPs are much narrower than the integrated profile, and their
positions within the integrated profile are stable.

The absence in our observations of GPs exceeding in intensity the
average pulse by more than a factor of 120, which observed for the pulsars PSR
B0531+21 and PSR B1937+21, may be attributable to the relatively
small (because of the long period of the pulsar PSR B0031--07)
number of observed periods (11480 compared to more than $10^6$
periods for PSR B0531+21 and PSR B1937+21). For these pulsars,
pulses exceeding in intensity the average level by more than a factor of 120
are observed once per approximately $10^5$ pulses.

Note that the pulsar PSR B0031--07, like PSR B1112+50 that we
detected previously (Ershov and Kuzmin 2003), does not belong to
the group of pulsars with the strongest magnetic fields on the
light cylinder ($B_{LC} \simeq 10^6$ G), which are considered to
be most promising for the search of GPs. For the pulsars PSR
B0031--07 and PSR B1112+50, the magnetic fields on the light
cylinder $B_{LC}$ are $\simeq$ 5 G. This may be a different class
of GPs.

The observed GP localization within a small part of the integrated
profile suggests that the GP radiation beam is much narrower than
the beam of the integrated profile. Hence, the probability to "see"
a pulsar GP is much lower than the probability to see normal,
broader pulses. Therefore, the number of observed pulsars with GPs
must be much smaller than the number of observed normal pulsars.

\section{CONCLUSIONS}
We have detected GPs from the pulsar PSR B0031--07. At 111 MHz,
the peak flux density of the strongest pulse is about 500 Jy,
which is more than a factor of 100 higher than the peak flux
density of the average pulse. Pulses for which the peak intensity
exceeds the peak intensity of the average pulse by more than a
factor of 50 are encountered approximately once per 300 periods.
The GPs are a factor of about 20 narrower than the integrated
profile and are clustered in the middle part of the integrated
profile. The brightness temperature of the observed GPs is $T_{B}
\geq 10^{26}$ K.

\begin{acknowledgments}
We wish to thank V.~V. Ivanova, K.~A. Lapaev, V.~D. Pugachev, and
A.~S. Aleksandrov for the software and technical support of the
observations. This work was supported in part by the Russian
Foundation for Basic Research (project no. 01-02- 16326) and
Program of the Presidium of the Russian Academy of Sciences
("Study of the Nature of Nonstationary Phenomena in Astrophysical
Objects in Various Ranges of the Electromagnetic Spectrum").

\end{acknowledgments}
\newpage
%
%

%
\newpage
\section{FIGURES CAPTIONS}

Fig. 1. Example of a GP from the pulsar PSR B0031--07.

Fig. 2. (a) Phases of the 43 observed GPs with peak flux densities
exceeding the peak flux density of the average pulse by more than
a factor of 50. (b) The strongest GP (1) together with the
integrated profile (2), which was magnified by a factor of 100.

\newpage
\begin{figure}
\setcaptionmargin{5mm}
\includegraphics{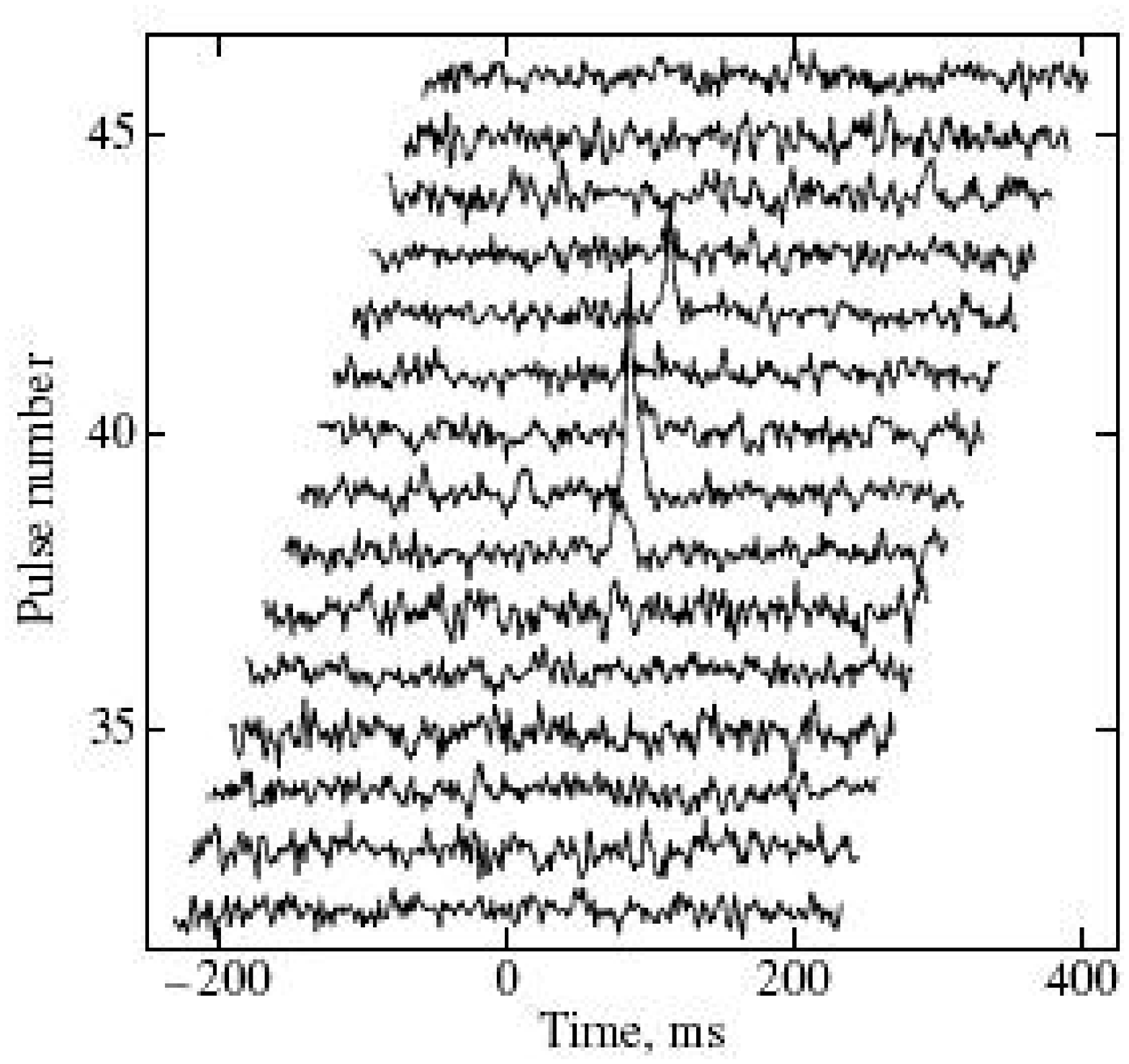} 
\end{figure}
\newpage
\begin{figure}
\setcaptionmargin{5mm}
\includegraphics{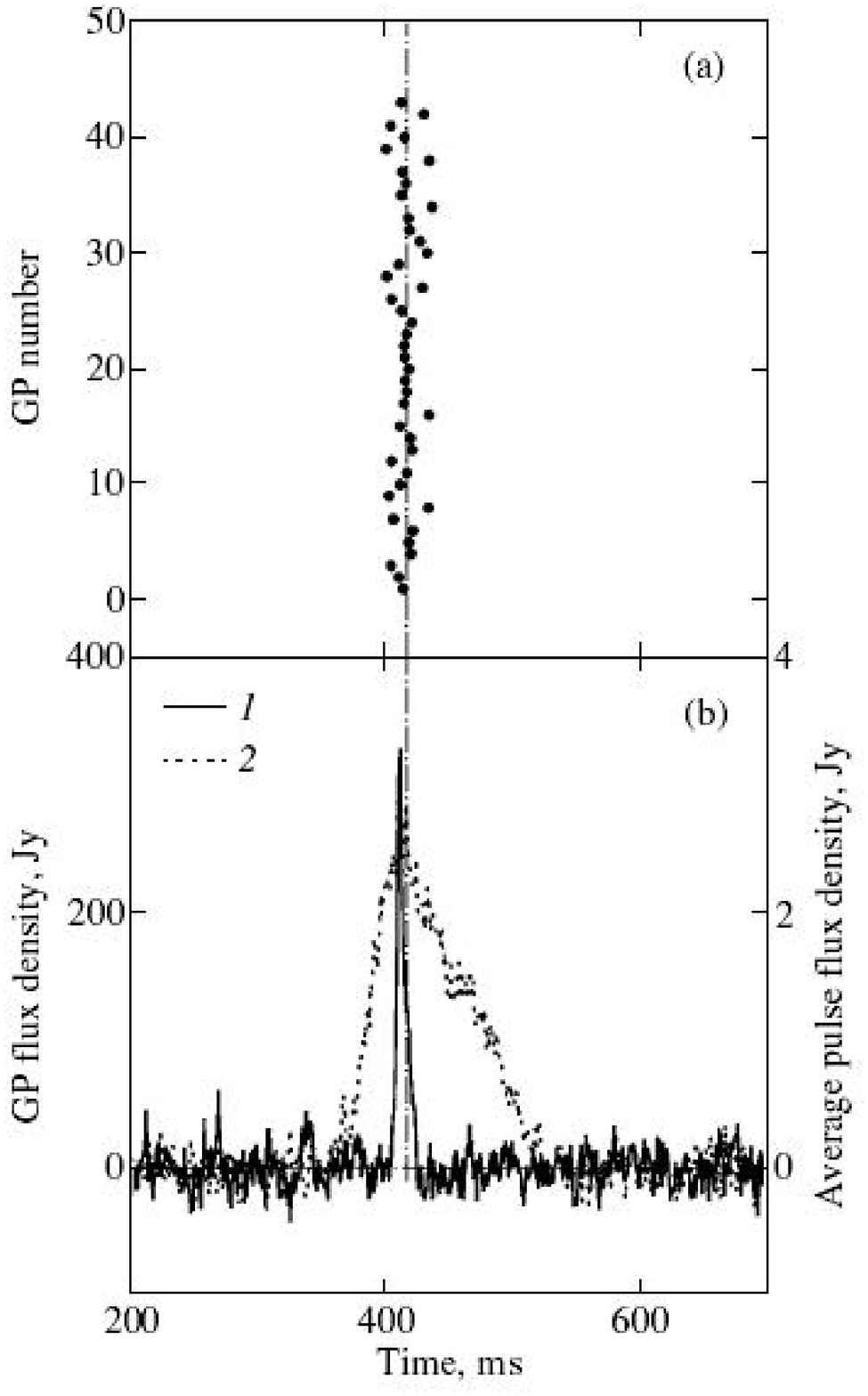} 
\end{figure}

\begin{thebibliography}{99}
\bibitem{1}
\refitem{article}
J. Cognard, J. A. Shrauner, J. H. Taylor, and S.
E. Thorsett, Astrophys. J. Lett. 457, L81 (1996).
\bibitem {2}
\refitem{article;rusjou}
A. A. Ershov and A. D. Kuzmin, Pis'ma
Astron. Zh. 29, 111 (2003) [Astron. Lett. 29, 91 (2003)].
\bibitem{3}
\refitem{article}
T. H. Hankins, J. S. Kern, J. C. Weatherall, and
J. A. Eilek, Nature 422, 141 (2003).
\bibitem{4}
\refitem{article}
V. A. Izvekova, A. D. Kuzmin, V. M. Malofeev, and
Yu. P. Shitov, Astrophys.Space.Sci. 78, 45 (1981).
\bibitem{5}
\refitem{article}
S. Johnston and R.W. Romani, Mon. Not. R. Astron.
Soc. 332, 109 (2002).
\bibitem{6}
\refitem{article}
S. Johnston and R. W. Romani, Astrophys. J. Lett.
590, L95 (2003).
\bibitem{7}
\refitem{article;rusjou}
S. V. Kostyuk, I. V. Kondrat'ev, A. D.
Kuzmin, et al., Pis'ma Astron. Zh. 29, 440 (2003) [Astron. Lett.
29, 387 (2003)].
\bibitem{8}
\refitem{article;rusjou}
A. D. Kuzmin and B. Ya. Losovsky, Pis'ma
Astron. Zh. 28, 25 (2002) [Astron. Lett. 28, 21 (2002)].
\bibitem{9}
\refitem{article}
S. C. Lundgren, J. M. Cordes, M. Ulmer, et al.,
Astrophys. J. 453, 433 (1995).
\bibitem{10}
\refitem{article} M. A. McLaughlin and J. M. Cordes, Astrophys. J.
596, 982 (2003).
\bibitem{11}
\refitem{article}
S. Phinney and J. H. Taylor, Nature 277, 117
(1979).
\bibitem{12}
\refitem{article}
R. W. Romani and S. Johnston, Astrophys. J. Lett.
557, L97 (2001).
\bibitem{13}
\refitem{article}
D. H. Staelin and J. M. Sutton, Nature 226, 69
(1970).
\bibitem{14}
\refitem{article;rusjou}
V. I. Shishov, V. M. Malofeev, A. V.
Pynzar', and T. V. Smirnova, Astron. Zh. 72, 485 (1995) [Astron.
Rep. 39, 428 (1995)].
\bibitem{15}
\refitem{misc}
J. H. Taylor, R. N. Manchester, A. G. Lyne, et al.,
Catalog of 706 Pulsars, (1995, Unpublished work).
\bibitem{16}
\refitem{book}
A. Wolszczan, J. M. Cordes, and D. R. Stinebring,
Millisecond Pulsars, Ed. by S. P. Reynolds and D. R. Stinebring
(NRAO, Green Bank, 1984), p. 63.
%
\end{thebibliography}
\end{document}